\documentclass{llncs}
\usepackage{makeidx}
\def\dom{{\mathsf{dom}}}
\def\Irr#1{{\mathsf{Irr}}(#1)}
\def\Pref#1{{\mathsf{Pref}}(#1)}
\def\IrrPref#1{{\mathsf{IrrPref}}(#1)}

\def\lstd#1{{\mathsf{lstdecomp}}(#1)}
\def\lst#1{{\mathsf{lstd}}(#1)}

\def\Ass{{\mathsf{Ass}}}
\def\ev{{\mathsf{ev}}}

\begin{document}
\mainmatter        
\pagestyle{headings}

\title{Partial monoids: associativity and confluence}

\author{Laurent Poinsot\inst{1}, G\'erard H.E. Duchamp\inst{1} and Christophe Tollu\inst{1}}

\institute{Universit\'{e} Paris-Nord, Institut Galil\'ee,\\
Laboratoire d'Informatique de Paris-Nord, UMR CNRS 7030,\\
F-93430 Villetaneuse, France}

\maketitle

\begin{abstract}
A partial monoid $P$ is a set with a partial multiplication $\times$ (and total identity $1_P$) which satisfies some  associativity axiom. The partial monoid $P$ may be embedded in the free monoid $P^*$ and the product $\times$ is simulated by a string rewriting system on $P^*$ that consists in evaluating the concatenation of two letters as a product in $P$, when it is defined, and a letter $1_P$ as the empty word $\epsilon$. In this paper we study the profound relations between confluence for such a system and associativity of the multiplication. Moreover we develop a reduction strategy to ensure confluence and which allows us to define a multiplication on  normal forms associative up to a given congruence of $P^*$. Finally we show that this operation is associative if, and only if, the rewriting system under consideration is confluent. 

{\bf{Key-words:}} Partial monoid, string rewriting system, normal form, associativity and confluence
\end{abstract}

\section{Introduction}
A partial monoid is a set equipped with a partially-defined multiplication, say $\times$, which is associative in the sense  that $(x\times y)\times z=x\times (y\times z)$ means that the left-hand side is defined if, and only if, the right-hand side is defined, and in this situation they are equal. A partial monoid is also assumed to have an identity element. Our original interest on such structures is due to the fact that they provide an algebraic framework for an abstract notion of connected components and the treatment of the exponential formula~\cite{DPGP09}. 

However another interesting feature of partial monoids motivates our work: their interpretation as a model of computation with errors. Programs can be interpreted as partial functions and their composition, when defined, simulate a sequential process. Abstracting this situation by considering programs as elements of a partial monoid, the notion of error occurs naturally: an error is nothing but the evaluation of a not defined product.  In order to locate the fault, we can set undefined products to be equal to some new symbol (an error flag), for instance $0$, \emph{i.e.}, $x\times y=0$ when $x\times y$ is undefined. Now, if we interpret an $n$-fold product $x_1\times x_2\times \cdots \times x_n$ as some sequential program, then if the evaluation of one of the factors is an error, the program itself is erroneous, in other terms, $0\times x=0=x\times 0$ for every $x$. This situation is not fully  satisfactory for the reason that the factor whose evaluation causes the error is lost by this crunch to zero. To fix this weakness, let us consider that the machine, which performs the execution $x_1\times x_2\times  \cdots \times x_n$, evaluates a factor $x_i\times x_{i+1}$ only when it is defined. In other terms, the machine only deals with error-free factors. The result of such an execution is a `` word '' $u_1\times \cdots \times u_k$ which may be seen as an exception handling: each factor $u_i$ marks faultless computations, while a product $u_j\times u_{j+1}$ labels an error. Obviously a word reduced to a single element represents the result of a program with no error at all.

Mathematically speaking, the previous situation is perfectly described first by embedding the partial monoid $P$ into the free monoid $P^*$ of words over the alphabet $P$, and second, by mimicking the execution of a program $w\in P^*$ as applications of the rewriting rules: if $w=uxyv$ and $x\times y$ is defined in $P$, then $w \Rightarrow u(x\times y)v$, and if $w=u1_Pv$ ($1_P$ is the identity of $P$), then $w\Rightarrow uv$. Actually an execution as described above is represented by reductions of the word as far as it is possible. In other words, an execution computes -- when it exists -- the normal form of the program $w$. This string rewriting system -- called a \emph{semi-Thue system} -- is easily seen to be terminating, \emph{i.e.} without infinite executions, property which guarantees existence, but not uniqueness, of normal forms. Seen as the result of an execution, a normal form should be unique. This is possible when the semi-Thue system is \emph{confluent}. 

The main objective of this work is to highlight the profound links between associativity and confluence  for such rewriting systems, that is, to give characterizations of confluence in terms of associativity, and \emph{vice versa}. In this paper, we exhibit the exact property the partial monoids must satisfy to ensure confluence of the system. Since this particular property does not hold in every partial monoid, we develop a strategy of reduction, called the \emph{left standard reduction}, which provides a unique normal form which is also a normal form for the initial system. Finally, using the left standard reduction, we equip the set of all normal forms with a total binary operation which is shown to be associative up to some monoidal congruence. In order to prove this result, we use another rewriting system on nonassociative words -- which allows us to move pairs of brackets to perform associativity -- in a way similar to the treatment of the coherence theorem for monoidal category~\cite{McL98}. Finally we show that the operation on normal forms is associative if, and only if, the semi-Thue system under consideration is confluent.   
\begin{note}
Most of the proofs of lemmas will be omitted, since they are free of technical difficulties. 
\end{note}
%
\section{Partial monoids}

A \emph{partial monoid} (see~\cite{DPGP09,Seg73,Wil95}) -- also sometimes called \emph{premonoid}~\cite{Bes03,BDM02} --  is a nonvoid set $P$ together with a partially-defined function $\times : P\times P \rightarrow P$, with domain of definition ${\mathsf{\dom}}(\times)\subseteq P\times P$, and a distinguished element, $1_P\in P$, called the \emph{identity}, such that
\begin{enumerate}
\item for every $x\in P$, $(x,1_P)$ and $(1_P,x)$ belong to $\dom(\times)$, and, $x\times 1_P=x=1_P\times x$;
\item for every $x,y,z\in P$, $(x,y)\in \dom(\times)$, $(x\times y,z)\in \dom(\times)$ if, and only if, $(y,z)\in\dom(\times)$, 
$(x,y\times z)\in \dom(\times)$, and, in both cases, $(x\times y)\times z=x\times (y\times z)$. 
\end{enumerate}
Let us consider the set $P^0=P\cup\{0\}$ obtained from $P$ by the adjunction of a new element $0$. The operation $\times$ is extended to the whole Cartesian product $P^0\times P^0$ as an operation $\times^0$ by setting $x\times^0 y = x\times y$ for every $(x,y)\in\dom(\times)$ and $x\times^0 y=0$ for remaining pairs of elements of $P^0$. This new structure is  a monoid (see~\cite{Ljap:eve}). From this we deduce that given $(x_1,\cdots,x_n)\in P^n$, if the $n$-fold product is defined for a particular choice of brackets, then it is defined for all bracketings, and the values are equal.   
\begin{example}\label{premierexemple}
\begin{enumerate}
\item Let $X$ be any set, and $2^X$ be the set of its subsets. We endow $2^X$ with the \emph{disjoint union} defined only for non-intersecting subsets. Then, $2^X$ is a partial monoid with $\emptyset$ as identity. Such monoids are useful to define a general setting for the exponential formula of combinatorics~\cite{DPGP09}.
\item\label{secondexempledupremierexemple} Let us consider the set 
\begin{equation}
P=\{\epsilon,a,b,c,ab,ac,ba,bc,ca,cb,abc,acb,bac,bca,cab,cba\}
\end{equation}
with the product $\times$ being concatenation of two words without common letters. Then $P$ is a partial monoid with the empty word $\epsilon$ as its identity. 
\end{enumerate}
\end{example}
\section{Basics on rewriting rules and normal forms}

\subsection{Abstract rewriting systems}

An \emph{abstract rewriting system} (see~\cite{BN99,Ter03} for more details) is a pair $(S,\Rightarrow)$ where $S$ is a set and $\Rightarrow$ is a binary relation on $S$, called \emph{one-step rewriting} or \emph{reduction relation}. If $(a,b)\in\ \Rightarrow$, then we write $a\Rightarrow b$ (`` \emph{$a$ is reduced by $\Rightarrow$ to $b$} '' and $a$ is said to be  \emph{reducible}).  The reflexive-transitive closure $\Rightarrow^*$ of $\Rightarrow$ is called the \emph{many-step rewriting relation} generated by $\Rightarrow$, while its symmetric-reflexive-transitive closure $\Leftrightarrow^*$, \emph{i.e.}, the equivalence relation generated by $\Rightarrow$, is called the \emph{convertibility relation} (generated by $\Rightarrow$). An abstract rewriting system is said to be 
\begin{enumerate}
\item \emph{terminating} if, and only if, $\Rightarrow$ is Noetherian;
\item \emph{confluent} if, and only if, for every $a,b,c\in S$ such that $a\Rightarrow^*b$ and $a\Rightarrow^*c$, there is $d$ such that $b\Rightarrow^* d$ and $c\Rightarrow^* d$;
\item \emph{locally confluent} if, and only if, for every $a,b,c\in S$ such that $a\Rightarrow b$ and $a\Rightarrow c$, there is $d$ such that $b\Rightarrow^* d$ and $c\Rightarrow^* d$.
\end{enumerate}  
If $a\in S$ is minimal with respect to $\Rightarrow$, \emph{i.e.}, there is no $b$ such that $a\Rightarrow b$, then $a$ is called a $\Rightarrow$-\emph{normal form} or, simply, a \emph{normal form}, or $a$ is said \emph{irreducible} (with respect to $\Rightarrow$). The set of all irreducible elements of $S$ is denoted $\Irr{S,\Rightarrow}$ or simply $\Irr{S}$ or $\Irr{\Rightarrow}$. If $a\in S$ and $b\in \Irr{S}$ such that $a\Rightarrow^* b$, then $b$ is called a \emph{normal form of $a$}. In a terminating  abstract rewriting system, every element has at least one normal form, and in a confluent abstract rewriting system the normal form of any element, if it exists, is unique \cite{Hue80}. 
\begin{lemma}[(Newman's lemma~\cite{Hue80,New42,Ter03})]
A terminating abstract rewriting system is confluent if, and only if, it is locally confluent. 
\end{lemma}
Therefore in a terminating and confluent abstract rewriting system, every element has a unique normal form. 

\subsection{Semi-Thue system}\label{semiThuesystem}

See~\cite{BO93,Jan88} for more details on string rewriting, and~\cite{Berg} for rewriting systems over algebraic structures.  
Let $X$ be any set. A \emph{semi-Thue system} $R$ on $X$ is a binary relation on $X^*$. An element of $R$ is called a(n) 
\emph{(elementary) rule}. The \emph{(single-step) reduction relation} on $X^*$ induced by the rules of $R$ is defined as follows: $uav \Rightarrow_R ubv$ whenever $u,v\in P^*$ and $(a,b)\in R$. Thus 
$(X^*,\Rightarrow_R)$ is an abstract rewriting system on $X^*$. We say that $R$ is locally confluent (resp. confluent, terminating) if the corresponding property holds for the abstract rewriting system $(X^*,\Rightarrow_R)$. We use $\Irr{X}$ or $\Irr{R}$ to denote $\Irr{X^*,\Rightarrow_R}$.  The reflexive-transitive closure $\Rightarrow_R^*$ of $\Rightarrow_R$ is called the \emph{reduction rule generated by $R$}. It can be seen as the smallest quasi-order relation containing $R$ which is compatible with concatenation (\cite{Laf07}). The convertibility relation $\Leftrightarrow_R^*$ (generated by $\Rightarrow_R$) is nothing else than the congruence generated by $R$, and called the \emph{Thue congruence} induced (or generated) by $R$. 
A pair $(u,v)\in X^*\times X^*$ is called a \emph{critical pair (of $R$)}  if, and only if, $u,v$ have either the form $u=u_1r_1$, $v=r_2v_2$ for some $u_1,v_1\in X^*$, $(\ell_1,r_1),(\ell_2,r_2)\in R$, $u_1\ell_1=\ell_2v_2$ and $|u_1|<|\ell_2|$  ($|w|$ is the length of a word $w$), or $u=r_1$, $v=v_1r_2v_2$ for some $v_1,v_2\in X^*$, 
$(\ell_1,r_1)$, $(\ell_2,r_2)$ and $\ell_1=v_1\ell_2v_2$. A critical pair of the first kind is called an \emph{overlap ambiguity}, while a critical pair of the second kind is an \emph{inclusion ambiguity}. 
A critical pair $(u,v)$ is \emph{convergent} if there is $w\in X^*$ such that $u\Rightarrow_R^* w$ and $v\Rightarrow_R^* w$. A critical pair $(u,v)$ such that $u=v$ is called \emph{trivial}. 
If a Thue system is known to be terminating, then local confluence -- and hence confluence -- holds if, and only if, each critical pair is convergent~\cite{Hue80} (actually this is a more general result that holds for term rewriting systems).

\section{Semi-Thue system associated with a partial monoid}

\subsection{First definitions}

Given a partial monoid $P$. Let $i_P:P\hookrightarrow P^*$ be the natural injection. Any element of $P^*$ may be written in a unique way as a word $i_P(x_1)\cdots i_P(x_n)$ for some $n\in \bbbn$ and $x_i\in P$ ($n=0$ leads to the empty word $\epsilon$). Moreover we sometimes use the notation $u=u_1\cdots u_n$ with the meaning that $u_i=i_P(x_i)$. We define the following semi-Thue system $R_P = \{(i_P(x)i_P(y),i_P(x\times y)) : (x,y)\in\dom(\times)\}\cup\{(i_P(1_P),\epsilon)\}$, call it \emph{the semi-Thue system associated with $P$}, which is easily seen to be terminating. A similar idea has been used in~\cite{Baer49,Bes03,BDM02,Bru58,Tam71} (see also~\cite{Smi51}). In what follows, when it is possible $R_P$ is denoted by $R$.  The set of irreducible elements $\Irr{P}$ with respect to $\Rightarrow_R$ is equal to  
\begin{equation}
\begin{array}{lll}
\{i_P(x_1)\cdots i_P(x_n) &:& \forall i,\ 1\leq i\leq n,\ x_i\in P\setminus\{1_P\}\ ,\\
 &&\forall i,\ 1\leq i < n,\ (x_i,x_{i+1})\not\in\dom(\times)\}\ .
\end{array}
\end{equation}
In particular it contains the empty word $\epsilon$ obtained for $n=0$, and every element of $P\setminus\{1_P\}$ (under the form $i_P(x)$). In case $P$ is a (total) monoid, then $\Irr{P}=i_P(P\setminus\{1_P\})\cup\{\epsilon\}$. 
\begin{note}
Since each $u\in \Irr{P}\setminus\{\epsilon\}$ belongs to $P^*$, then $u$ has a unique decomposition of the form $i_P(x_1)\cdots i_P(x_n)$, $x_i\in P\setminus\{1_P\}$, $1\leq i\leq n$, $(x_i,x_{i+1})\not\in\dom(\times)$, $1\leq i<n$. 
\end{note}
Note that $\Irr{P}$ is prefix-closed\footnote{It is also closed under factors~\cite{BPR09}.}. Recall that $u$ is a prefix of $v$ if, and only if, there is $u' \in P^*$ such that $v=uu'$. Let $u\leq_P v$ be the relation `` $u$ is a prefix of $v$ ''. This partial order relation on $P^*$ satisfies $u\leq_P v$ and $u'\leq_P v$ implies that $u$ and $u'$ are comparable, \emph{i.e.}, $u\leq_P u'$ or $u'\leq_P u$ (see~\cite{BPR09}). In what follows, $\Pref{w}$ denotes the set 
$\{u\in P^* : u\leq_P w\}$ of all prefixes of $w$, totally ordered by the restriction of $\leq_P$. 

\subsection{Discussion about the confluence}\label{subsect:discussion}

Let $P$ be a partial monoid and $R$ be its associated semi-Thue system. 
In general, $R$ is not confluent (since it is not locally confluent). Indeed, the critical pair $(i_P(ab)i_P(a),i_P(a)i_P(ba))$ obtained from example~\ref{premierexemple}.\ref{secondexempledupremierexemple} is not convergent. We call \emph{essential} any critical pair of the form $((i_P(a)i_P(z),i_P(x)i_P(b))$ such that there is some $y\in P$ with $(x,y)\in\dom(\times)$, $(y,z)\in\dom(\times)$, $x\times y = a$ and $y\times z=b$. 
\begin{lemma}\label{essentialcriticalpairs}
The semi-Thue system $R$ is confluent if, and only if, every essential critical pair converges.
\end{lemma}
An essential critical pair may be trivial (take $y=1_P$) so we try now to figure out those on which  local confluence relies.  
The set of all essential critical pairs may be decomposed into several subsets. Let $(u,v)=(i_P(a)i_P(z),i_P(x)i_P(b))$ be an essential critical pair which comes from an overlap ambiguity $i_P(x)i_P(y)i_P(z)$ with $(x,y)\in\dom(\times)$, $x\times y=a$ and $(y,z)\in\dom(\times)$, $y\times z=b$. We say that $(u,v)$ is of \emph{type (A)} if $(a,z)\not\in \dom(\times)$ (and therefore $(x,b)\not\in\dom(\times)$). The critical pair $(u,v)$ is of \emph{type (B)} if $(a,z)\in\dom(\times)$ (and therefore $(x,b)\in\dom(\times)$): such a critical pair is convergent. The two types are obviously disjoint and cover all the essential critical pairs. We also say that a critical pair $(u,v)=(i_P(a)i_P(z),i_P(x)i_P(b))$ of type (A) (so $(a,z)\not\in\dom(\times)$, $(x,b)\not\in\dom(\times)$) is of \emph{type (A1)} if $a=x$, $b=z$ (so in particular $(x,z)\not\in\dom(\times)$); we immediately notice that a critical pair of type (A1) is trivial. A pair of type (A) is said to be  of \emph{type (A0)} if $u=i_P(a)i_P(z)$, $v=i_P(x)i_P(b)$, and $a\not=x$ or $b\not=z$. Types (A0) and (A1) are disjoint (in the second case $u=v$ while in the first one $u\not=v$). Each essential critical pair of type (A) is either of type (A0) or of type (A1). 
\begin{lemma}\label{lemmeconfluenceessentialcpA0}
The semi-Thue system $R$ is confluent if, and only if, there is no critical pair of type (A0), or equivalently, if, and only if, each essential critical pair of type (A) is of type (A1).
\end{lemma}
\begin{proof}
The above discussion shows that the only possible non convergent essential critical pairs are of type (A0). Suppose that $(u,v)=(i_P(a)i_P(z),i_P(x)i_P(b))$ is an essential critical pair of type (A0), \emph{i.e.}, $(a,z)\not\in\dom(\times)$, 
$(x,b)\not\in\dom(\times)$, $x\not=a$ or $z\not=b$, and there is $y\in P$ with $(x,y)\in\dom(\times)$, $x\times y=a$, $(y,z)\in\dom(\times)$, $y\times z=b$,  $(\ell_1,r_1)=(i_P(y)i_P(z),i_P(y\times z))$, 
$(\ell_2,r_2)=(i_P(x)i_P(y),i_P(x\times y))$. From the assumptions we deduce that $x\not=1_P$ (otherwise $(y,z)\not\in\dom(\times)$), $z\not=1_P$ (otherwise $(x,y)\not\in\dom(\times)$) and $y\not=1_P$ (otherwise $x=a$ and $z=b$). Moreover $a=x\times y\not=1_P$ (otherwise $(1_P,z)=(x\times y,z)\not\in\dom(\times)$), $b=y\times z\not=1_P$ (otherwise $(x,1_P)=(x,y\times z)\not\in\dom(\times)$). So no rewriting rule can be applied on $u$ or on $v$. Since $u\not=v$ (by assumption), $(u,v)$ is not convergent.  Suppose that $R$ is confluent. So by lemma~\ref{essentialcriticalpairs}, every essential critical pair is convergent. But critical pairs of type (A0) cannot be convergent, so in this case, there is no such critical pair. \qed
\end{proof}
\begin{example}\label{example2}
Let $P=\{1,x,y,z\}$ be a set with four elements equipped with a product $\times$ for which the only non trivial pairs  (\emph{i.e.} pairs without occurrences of the identity $1$) in its domain are $(x,y),(y,y)$ and $(y,z)$. We suppose that $x\times y = x$, $y\times y=y$ and $y\times z=z$. Then $R$ is confluent because there is no critical pair of type (A0). 
\end{example}
Confluence is obtained for a rather important class of partial monoids. A partial monoid $P$ is called \emph{catenary associative} (see~\cite{Ljap:eve} for the definition of `` catenary associativity '' in a partial magma, which is adapted for our purpose; see also~\cite{Gud72}) if, and only if, for all $x,y,z \in P$, if $y\not=1_P$, $(x,y)\in\dom(\times)$ and $(y,z)\in\dom(\times)$, then $(x\times y,z)\in\dom(\times)$ (and also $(x,y\times z)\in \dom(\times)$ by associativity in any partial monoid). We need that $y\not=1_P$ otherwise the monoid would be total. None of the monoids of example~\ref{premierexemple} is catenary associative. Every (total) monoid is catenary associative. The set of arrows of a small category (see~\cite{McL98}) together with an adjoined total identity (and the obvious extension of composition) is a catenary associative partial monoid. It is easy to prove that in the catenary case there is no critical pair of type (A0).
\begin{lemma}\label{confluencepourcatenarity}
Let $P$ be a partial monoid. If $P$ is catenary associative, then the semi-Thue system $R$ is confluent.
\end{lemma}
Partial monoids from example~\ref{premierexemple} have non confluent associated semi-Thue systems while 
the monoid of example~\ref{example2} is not catenary but $R$ is confluent.
 
\subsection{Left standard reduction}\label{standardstrategyrewriting}

In order to get a unique normal form property, even for non confluent semi-Thue system $R$, we restrict $R$ by allowing only rewriting steps from `` left to right ''. This algorithm of reduction (informally described below) will ensure both termination and confluence, and therefore computes a unique normal form which is also a normal form for $R$. 
\begin{enumerate}
\item \texttt{Input:} a word $w\in P^*$.
\item \texttt{Erase} all occurrences of $i_P(1_P)$ \texttt{in} $w$. \texttt{Result} $w'\in (P\setminus\{1_P\})^*$.
\item \texttt{While} $w'\not\in \Irr{P}$ \texttt{do} let $r:=i_P(x)i_P(y)$ be the first factor of $w'$ (from left to right) such that $(x,y)\in \dom(\times)$. \texttt{If} $x\times y=1_P$, \texttt{then erase} $r$ from $w'$ \texttt{else substitute} $r$ by $i_P(x\times y)$ in $w'$. 
\item \texttt{Output:} $w'\in \Irr{P}$. 
\end{enumerate}
 
First of all let $R_1=\{(i_P(1_P),\epsilon)\}$. This semi-Thue system $R_1$ is terminating and confluent (since it has no critical pair). Thus every element of $P^*$ has a unique normal form in $\Irr{R_1}=(P\setminus\{1_P\})^*$. 
\begin{lemma}\label{lemmemaxirrpref}
Let $w\in (P\setminus\{1_P\})^*$. Then 
\begin{enumerate}
\item $\IrrPref{w}=\Irr{P}\cap\Pref{w}$ admits a maximum $w_m$ (for the total order $\leq_P$ restricted to $\IrrPref{w}$);
\item $w_m=w$ if, and only if, $w\in \Irr{P}$;
\item If $w\not\in \Irr{P}$, then there is a unique $4$-tuple $(u,x,y,v)\in (P\setminus\{1_P\})^*\times (P\setminus\{1_P\})\times(P\setminus\{1_P\})\times(P\setminus\{1_P\})^*$ such that 
\begin{enumerate}
\item $w_m=ui_P(x)$;
\item $w=ui_P(x)i_P(y)v$;
\item $(x,y)\in\dom(\times)$.
\end{enumerate}
\end{enumerate}
\end{lemma}
\begin{proof}
First of all, $\IrrPref{w}$ is nonvoid because $\epsilon\in \IrrPref{w}$. Since $\IrrPref{w}$ is a subset of $\Pref{w}$ and as such is totally ordered by the restriction of $\leq_P$, it is sufficient to show that $\IrrPref{w}$ admits a maximal element, that is, an element $w_m\in\IrrPref{w}$ such that there is no $w'\in\IrrPref{w}$ with $w\leq_P w'$ and $w'\not=w$. 
\begin{itemize}
\item Suppose that $w\not\in\Irr{P}$. Since $w\in\Irr{R_1}$, that means that $|w|>1$ and there is at least one integer $i$, 
$1\leq i< |w|$ such that $w_i=i_P(x)$, $w_{i+1}=i_P(y)$, $(x,y)\in\dom(\times)$ and $x\not=1_P$, $y\not=1_P$. Let $i_0$ be the least such integer. Let $w_m=w_1\cdots w_{i_0}$. Then, by definition of $i_0$, $w_m\in \IrrPref{w}$. Let $w'\in \IrrPref{w}$ such that $w_m\leq_P w'$. Then either $w_m=w'$ or $w'_{i_0}w'_{i_0+1}$ may be rewritten but in the latter case, $w'\not\in \Irr{P}$.  
\item Suppose that $w\in \Irr{P}$. In this case, $w_m=w$. So we are done with (1). Note that the converse is obvious, and (2) holds.
\item Concerning (3), let $w_{i_0}=i_P(x)$, $w_{i_0+1}=i_P(y)$ (with $x\not=1_P$, $y\not=1_P$ and $(x,y)\in\dom(\times)$). Let $u=w_1\cdots w_{i_0-1}$ (thus $u=\epsilon$ if, and only if, $i_0=1$). Then $w_m=ui_P(x)$. Moreover $w=w_mw_{i_0+1}\cdots w_{|w|}=ui_P(x)i_P(y)v$ where $v=w_{i_0+2}\cdots w_{|w|}$ (thus $v=\epsilon$ if, and only if, $i_0+1=|w|$). \qed
\end{itemize} 
\end{proof}
For $w \in (P\setminus\{1_P\})^*\setminus\Irr{P}$, the $4$-tuple $(u,x,y,v)$ of lemma~\ref{lemmemaxirrpref} is called the \emph{left-standard decomposition of $w$}, and denoted by $\lstd{w}$. 
\begin{lemma}\label{invertiblelements}
Let $x\in P$ be a right (resp. left) invertible element. Then for every $y\in P$, $(y,x)\in \dom(\times)$ (resp. $(y,x)\in\dom(\times)$). In particular, if $x$ is invertible, then every pair $(x,y)$ and $(y,x)$ belong to $\dom(\times)$
\end{lemma}
\begin{proof}
Suppose that $x\in P$ is right (resp. left) invertible. Let $x'\in P$ such that $(x,x')\in\dom(\times)$ and $x\times x'=1_P$ (resp. $(x',x)\in\dom(\times)$ and $x'\times x=1_P$). Let $y\in P$ such that $(y,x)\not\in \dom(\times)$ (resp. $(x,y)\not\in \dom(\times)$). But $(y,x\times x')=(y,1_P)\in\dom(\times)$ (resp. $(x'\times x,y)=(1_P,y)\in\dom(\times)$) and therefore, by associativity in $P$, $(y,x)\in\dom(\times)$ (resp. $(x,y)\in\dom(\times)$), that is, a contradiction. The last assertion of the lemma is straightforward. \qed
\end{proof}
\begin{lemma}\label{suiteinvertiblelements}
Let $u\in\Irr{P}\setminus\{\epsilon\}$ such that there is some $i\in \bbbn$, $1\leq i\leq |u|$ with $u_i=i_P(x)$ and $x$ is right-invertible (resp. left-invertible). Then $i=1$ (resp. $i=|u|$). In particular, if $x$ is invertible, then $u=i_P(x)$. 
\end{lemma}
\begin{proof}
Suppose that $u_i=i_P(x)$ such that $x$ is right (resp. left) invertible. According to lemma~\ref{invertiblelements}, for every $y\in P$, $(y,x)\in\dom(\times)$ (resp. $(x,y)\in\dom(\times)$). Now suppose that $i\not=1$ (resp. $i\not=|u|$). Let $u_{i-1}=i_P(y)$ (resp. $u_{i+1}=i_P(y)$). Because $u$ is irreducible, we have the contradiction $(y,x)\not\in\dom(\times)$ (resp. $(x,y)\not\in\dom(\times)$). The last assertion is trivial. 
\qed
\end{proof}
\begin{lemma}
Let $w \in (P\setminus\{1_P\})^*\setminus\Irr{P}$. Let $\lstd{w}=(u,x,y,v)$. If $x\times y=1_P$, then $u=\epsilon$, and, in particular, $w_m=i_P(x)$ (and therefore $\IrrPref{w}=\{\epsilon,i_P(x)\}$) and $\lstd{w}$ has the form $(\epsilon,x,y,v)$. 
\end{lemma}
\begin{proof}
Suppose that $x\times y=1_P$. Then, according to lemma~\ref{invertiblelements}, for every $z\in P$, $(z,x)\in \dom(\times)$ and $(y,z)\in\dom(\times)$. 
Now we can deduce that, since $ui_P(x)=w_m\in\Irr{P}$, then $u=\epsilon$ according to lemma~\ref{suiteinvertiblelements}.    
\qed
\end{proof}
\begin{lemma}
Let $A=\{w\in (P\setminus\{1_P\})^*\setminus\Irr{P} : \lstd{w}=(\epsilon,x,y,v),\ x\times y=1_P\}$ and 
$B=\{w\in (P\setminus\{1_P\})^*\setminus\Irr{P} : \lstd{w}=(u,x,y,v),\ x\times y\not=1_P\}$. Then $A\cap B=\emptyset$ and 
$A\cup B = (P\setminus\{1_P\})^*\setminus\Irr{P}$.
\end{lemma}
Now let define $\rho_A = \{(w,v)\in A\times(P\setminus\{1_P\})^* : \lstd{w}=(\epsilon, x,y,v)\}$ and 
$\rho_B = \{(w,w')\in B\times (P\setminus\{1_P\})^* : \lstd{w}=(u,x,y,v),\ w'=ui_P(x\times y)v\}\ .$
 Both binary relations are functional (that is, $(x,y),(x,y')\in \rho_C$ implies that $y=y'$ for $C=A,B$).  We write $\rho_C(w)=v$ for $(w,v)\in \rho_C$ ($C\in \{A,B\}$), in such a way that $\rho_A : A \rightarrow (P\setminus\{1_P\})^*$ and $\rho_B : B\rightarrow(P\setminus\{1_P\})^*$. It is not difficult to see that $\rho_A\cup\rho_B$ is a functional relation and a locally confluent  abstract rewriting system on $(P\setminus\{1_P\})^*$ which is also terminating, and thus confluent. Moreover its set of normal forms is exactly $\Irr{P}$.

Let us consider the abstract rewriting system on $P^{*}$, called \emph{left standard reduction}, 
\begin{equation}
\lst{R} =\ \Rightarrow_{R_1} \cup\ \rho_A \cup \rho_B\ .
\end{equation}
The abstract rewriting system $\lst{R}$ is terminating since the length of a word is reduced by any one-step reduction. We can also easily check that it is locally confluent, and therefore confluent. The set of irreducible elements with respect to $\lst{R}$ is $\Irr{P}$. 
\begin{note}\label{quelquesremarques}
The many-step rewriting rule $\Rightarrow_{\lst{R}}^*$ generated by $\lst{R}$  and the equivalence relation $\Leftrightarrow_{\lst{R}}^*$ generated by $\lst{R}$ are respectively included in $\Rightarrow_R^*$ and $\Leftrightarrow_{R}^*$ (to prove this it is sufficient to see that $\lst{R}\subseteq \Rightarrow_R^*$). 
\end{note}
Since $\lst{R}$ is terminating and confluent, for every $w \in P^*$, there is one and only one $w'\in \Irr{R}$ such that $(w,w')\in \lst{R}^*$. Let ${\mathsf{lstd}}:P^*\rightarrow \Irr{P}$ be the mapping that maps a word to its normal form by $\lst{R}$-reductions only. 
\begin{lemma}
Let $u,v,w\in P^*$ such that $(u,v)\in \lst{R}$. Then $(uw,vw)\in\lst{R}$. 
\end{lemma}
\begin{proof}
Suppose that there is at least one $i$ such that $u_i=i_P(1_P)$, then only the reduction relation $\Rightarrow_{R_1}$ may be applied. In particular $v$ is obtained by erasing (exactly) one occurrence of $i_P(1_P)$ from $u$, saying $u_i$. Therefore $vw$ is obtained by erasing the same occurrence $u_i$ in the prefix $u$ of $uw$. Suppose that $w\in A\cup B$. If $w \in A$, then $\lstd{u}=(\epsilon,x,y,v)$ and $v=\rho_A(u)$. Now $uw\in A$ and $\lstd{uw}=(\epsilon,x,y,vw)$ in such a way that $\rho_A(uw)=vw$ as expected. Let suppose that $u\in B$. Let $\lstd{w}=(u',x,y,v')$ (with $x\times y\not=1_P$) in such a way that $v=u'i_P(x\times y)v'$. Then $uw\in B$ and $\lstd{uw}=(u',x,y,v'w)$, so $\rho_B(uw)=u'i_P(x\times y)v'w=vw$. 
\qed
\end{proof}

\begin{lemma}\label{rightcongruence}
Let $u,v,w\in P^*$ such that $u\Rightarrow_{\lst{R}}^* v$, \emph{i.e.}, $(u,v)$ belongs to the reflexive-transitive closure of $\lst{R}$. Then $uw\Rightarrow_{\lst{R}}^* vw$. 
\end{lemma}
\begin{proof}
Use the previous lemma several times. \qed
\end{proof}

\begin{lemma}\label{lempourrightmodule}
For every $u,v\in P^*$, $\lst{\lst{u}v}=\lst{uv}$.
\end{lemma}
\begin{proof}
By definition $u\Rightarrow_{\lst{R}}^* \lst{u}$. Therefore $uv \Rightarrow_{\lst{R}}^* \lst{u}v$ for any $v\in P^*$ according to the previous lemma. By uniqueness of the normal form, $\lst{uv}=\lst{\lst{u}v}$. 
\qed
\end{proof}
\begin{note}\label{equivalencelstdetRconfluent}
\begin{enumerate}
\item According to lemma~\ref{lempourrightmodule}, $\Irr{P}$ is a right $P$-module (see~\cite{Eil76}).
\item In general the symmetric-reflexive-transitive closure $\Leftrightarrow_{\lst{R}}^*$ of the left standard strategy $\lst{R}$ is only a right congruence of $P^*$. 
\item Let $R,S$ be two binary relations on some set $X$. We say that $R$ and $S$ are \emph{equivalent}, in symbol $R\equiv S$, if, and only if, $\Leftrightarrow_R^*=\Leftrightarrow_S^*$ (where $\Leftrightarrow_B^*$ is the equivalence relation generated by a binary relation $B$). Now suppose that $R$ is itself confluent, then $\lst{R}\equiv \Rightarrow_R$. 
\end{enumerate}
\end{note}

\section{Monoid-like structures on $\Irr{P}$}

No matter $R$ be confluent or not, we can always equip $\Irr{P}$ with a monoid-like structure. However in general this operation is only associative up to the congruence $\Leftrightarrow_R^*$. For every $(u,v)\in \Irr{P}^2$, let us define $u\star v = \lst{uv}$.  

In general, $\star$ is not associative. For instance, let $x,y,z \in P$ such that $(x,y)\in\dom(\times)$, $x\times y=a$, $(y,z)\in\dom(\times)$, $y\times z=b$, and $(i_P(a)i_P(z),i_P(x)i_P(b))$ is a critical pair of type (A0). Then $(i_P(x)\star i_P(y))\star i_P(z)=i_P(a)i_P(z)$, and $i_P(x)\star (i_P(y)\star i_P(z))=i_P(x)i_P(b)$. Thus $(i_P(x)\star i_P(y))\star i_P(z)\not=i_P(x)\star (i_P(y)\star i_P(z))$. 
\begin{lemma}\label{associativemoduloequivalence}
The operation $\star$ is `` associative modulo $\Leftrightarrow_R^*$ '', \emph{i.e.}, for all $u,v,w\in\Irr{P}$, 
$(u\star v)\star w\Leftrightarrow_R^* u\star (v\star w)$. 
\end{lemma}
\begin{proof}
On one side, 
\begin{equation}
\begin{array}{lll}
(u\star v)\star w&=&\lst{\lst{uv}w}\\
&=&\lst{(uv)w}\ \mbox{(according to lemma~\ref{lempourrightmodule})}\\
&=&\lst{u(vw)}\ ,
\end{array}
\end{equation}
on the other side, $u\star(v\star w)=\lst{u\lst{vw}}$. 
According to note~\ref{quelquesremarques}, $vw\Leftrightarrow_R^*\lst{vw}$ (since for any $x\in P^*$, $x\Rightarrow_{\lst{R}}^* \lst{x}$, which implies that $x\Rightarrow_{R}^* \lst{x}$, and therefore $x\Leftrightarrow_{R}^* \lst{x}$). Because $\Leftrightarrow_{R}^*$ is a congruence of $P^*$, $u(vw)\Leftrightarrow_R^*u\lst{vw}$. We conclude with the following sequence of equivalences.
\begin{equation}\label{sequencedequivalences}
\begin{array}{lll}
(u\star v)\star v&=&\lst{u(vw)}\\
&\Leftrightarrow_{R}^*& u(vw)\\
&\Leftrightarrow_{R}^*& u\lst{vw}\\
&\Leftrightarrow_R^*& \lst{u\lst{vw}}\\
&=&u\star(v\star w)\ .
\end{array}
\end{equation}
\qed
\end{proof}
Actually it is possible to prove that bracketings are irrelevant for $\star$ in the sense that any other choice of bracketings for the product $(\cdots((x_1\star x_2)\star x_3)\star\cdots )\star x_n)$ will evaluate to a normal form which is equivalent modulo the Thue congruence $\Leftrightarrow_R^*$. 
Let $X$ be a set and ${\mathsf{Mag}}(X)$ be the free magma generated by $X$ \cite{BouAlg}. This set is equipotent to the free $\Sigma$-algebra generated by $X$ with a unique symbol of function of arity $2$~\cite{Gra68}: the set of all binary trees with leaves in $X$. Every element of ${\mathsf{Mag}}(X)\setminus X$ may be written in a unique way as $t_1t_2$ ($t_1,t_2\in {\mathsf{Mag}}(X)$). Let ${\mathsf{Ass}}=\{((t_1t_2)t_3,t_1(t_2t_3)):t_1,t_2,t_3\in {\mathsf{Mag}}(X)\}$. We extend this binary relation to a term rewriting system $\Rightarrow_{\Ass}$ on ${\mathsf{Mag}}(X)$ in the usual way (see~\cite{BN99}) which allows us to rewrite a subtree of the form $(t_1 t_2)t_3$ in a given tree. This term rewriting system is terminating. To see that, it is sufficient to check that the \emph{rank}\footnote{Inspired from the rank of \cite{McL98} used  for the coherence theorem of monoidal categories.} ${\mathsf{rk}}:{\mathsf{Mag}}(X)\rightarrow\bbbn$ of a tree, defined by ${\mathsf{rk}}(x)=0$ for every $x \in X$ and ${\mathsf{rk}}(t_1 t_2)={\mathsf{rk}}(t_1)+{\mathsf{rk}}(t_2)+\ell(t_1)-1$ where $\ell(t)$ is the number of leafs of $t$ ($\ell(x)=1$ for every $x \in X$), strictly decreases at each application of a rewriting rule. Note that ${\mathsf{rk}}(t)=0$ if all closing brackets are in backside position. Moreover $\Ass$ is locally confluent: the only critical pairs (see~\cite{BN99} for a general notion of critical pairs for term rewriting systems, see also~\cite{GP00}) comes from an overlap of  
$((xy)z)w$ (this is basically due to the consideration of the most general unifier between the subterm $xy$ of the term $(xy)z$ and $(xy)z$ itself). So the critical pair is $((x(yz))w,(xy)(zw))$ given by two different applications of $\Rightarrow_{\Ass}$ on the tree $((xy)z)w$. This critical pair converges (it satisfies Stasheff's pentagon~\cite{Sta63}, made famous  in~\cite{McL98}). Since $\Rightarrow_{\Ass}$ is terminating, it is confluent, which is not amazing at all since the rule $(xy)z\rightarrow x(yz)$ provides a `` canonical system '' for the variety of semigroups~\cite{LeCh86}. As usually $\Rightarrow_{\Ass}^*$ denotes the reflexive-transitive closure of $\Rightarrow_{\Ass}$. Now, $(\Irr{P},\star)$ is also a magma. Let $\ev : {\mathsf{Mag}}(\Irr{P}) \rightarrow \Irr{P}$ be the unique homomorphic extension of the identity, called the \emph{morphism of evaluation} (see~\cite{Lal90} for the definition of such a morphism in any $\Sigma$-algebra). For every $x\in \Irr{P}$, $\ev(x)=x$ and $\ev(t_1t_2)=\ev(t_1)\star \ev(t_2)$. 
\begin{proposition}
Let $t_1,t_2\in{\mathsf{Mag}}(\Irr{P})$. If $t_1\Rightarrow_{\Ass}^* t_2$, then $\ev(t_1)\Leftrightarrow_{R}^* \ev(t_2)$. 
\end{proposition}
\begin{proof}
According to lemma~\ref{associativemoduloequivalence}, if $(t_1,t_2)\in\Ass$, then $\ev(t_1)\Leftrightarrow_{R}^*\ev(t_2)$. By structural induction on ${\mathsf{Mag}}(\Irr{P})$ we easily prove that if $t_1\Rightarrow_{\Ass} t_2$, then $\ev(t_1)\Leftrightarrow_R^*\ev(t_2)$. Finally, by transitivity of $\Leftrightarrow_{R}^*$, from $t_1\Rightarrow_{Ass}^* t_2$, we deduce that $\ev(t_1)\Leftrightarrow_R^* \ev(t_2)$ as expected. \qed
\end{proof}
Roughly speaking this result means that the order of the evaluation of $\star$ products is irrelevant with respect to the Thue congruence. We cannot expect more from a non confluent semi-Thue system $R$ (see proposition~\ref{strictassociativity}).
\begin{note}
A similar result may be obtained in a more general context: let $(M,*)$ be a magma and $\cong$ a congruence~\cite{BouAlg} on $M$. Suppose that for every $x,y,z\in M$, $(x*y)*z\cong x*(y*z)$. The following statement holds: for every $t,t'\in {\mathsf{Mag}}(M)$, if $t\Rightarrow_{\Ass}^* t'$, then $\ev(t)\cong \ev(t')$ (where $\ev : {\mathsf{Mag}}(M)\rightarrow M$ is the corresponding evaluation morphism).
\end{note}
\begin{proposition}\label{strictassociativity}
The operation $\star$ is associative if, and only if, $R$ is confluent. 
\end{proposition}
\begin{proof}
Suppose that $R$ is confluent. According to remark~\ref{equivalencelstdetRconfluent}, $\lst{R}\equiv \Rightarrow_R$, \emph{i.e.}, 
$\Leftrightarrow_{\lst{R}}^*=\Leftrightarrow_R^*$. Therefore we can replace each occurrence of $\Leftrightarrow_{R}^*$ by an occurrence of $\Leftrightarrow_{\lst{R}}^*$ in the sequence of equivalences~(\ref{sequencedequivalences}) of the proof of lemma~\ref{associativemoduloequivalence}. We obtain $(u\star v)\star w=\lst{uvw}\Leftrightarrow_{\lst{R}}^* \lst{u\lst{vw}}=u\star(v\star w)$. Since there is one and only normal form in each  equivalence class modulo $\Leftrightarrow_R^*$, we have $\lst{uvw}=\lst{u\lst{vw}}$, and thus $(u\star v)\star w=u\star (v\star w)$. Conversely, suppose that $\star$ is associative. Let $(i_P(a)i_P(z),i_P(x)i_P(b))$ be a critical pair of type (A0), that is, there is some $y\in P$ such that $(x,y)\in\dom(\times)$, $x\times y=a$, $(y,z)\in \dom(\times)$, $y\times z=b$, $(a,z)\not\in\dom(\times)$ and $x\not=a$ or $z\not=b$. Then $(i_P(x)\star i_P(y))\star i_P(z)=i_P(a)i_P(z)\not=i_P(x)i_P(b)=i_P(x)\star(i_P(y)\star i_P(z))$, which contradicts the assumption. Therefore there is no critical pair of type(A0), and by lemma~\ref{lemmeconfluenceessentialcpA0}, $R$ is confluent. 
\qed
\end{proof}
\begin{note}
Clearly, if $R$ is confluent, then $\Irr{P}$ is isomorphic to $P^*/\Leftrightarrow_{R}^*$. Moreover, if $P$ is a usual monoid, then $\phi:\Irr{P}=i_P(P\setminus\{1_P\})\rightarrow P$ defined by $\phi(\epsilon)=1_P$, and $\phi(i_P(x))=x$ is an isomorphism of monoids.
\end{note}

%

\begin{thebibliography}{}
%
\bibitem{BN99}
Baader, F., Nipkow, T.:
\emph{Term rewriting and all that}.
Cambridge University Press, 1999
%
\bibitem{Baer49}
Baer, R.:
Free Sums of Groups and Their Generalizations. An Analysis of the Associative Law. 
American Journal of Mathematics. \textbf{71}(3) (1949) 706--742
%
\bibitem{Bes03}
Bessis, D.:
The dual braid monoid.
Ann. Scient. \'Ec. Norm. Sup. $4^e$ s\'erie, t. 36 (2003) 647--683
%
\bibitem{BDM02}
Bessis, D., Digne, F., Michel J.:
Springer theory in braid groups and the Birman–Ko–Lee monoid.
Pacific J. Math. \textbf{205} (2002) 287-–310 
%
\bibitem{Berg}
Bergman, G.M.:
The diamond lemma for ring theory. 
Advances in Mathematics. \textbf{29} (1978) 178--218
%
\bibitem{BPR09}
Berstel, J., Perrin, D., Reutenauer, C.:
\emph{Codes and Automata},
volume 129 of \emph{Encyclopedia of Mathematics and its Applications}.
Cambridge University Press, 2009
%
\bibitem{BO93}
Book, R.V., Otto, F.:
\emph{String-rewriting systems}, 
\emph{Texts and Monographs in Computer Science}.
Springer-Verlag, 1993
%
\bibitem{BouAlg}
Bourbaki, N.:
\emph{Algebra - chapters 1-3}.
Springer, 1989
%
\bibitem{Bru58}
Bruck, R.H.:
\emph{A survey of binary systems}, 
Springer-Verlag, 1958
%
\bibitem{DPGP09}
Duchamp, G.H.E., Poinsot, L., Goodenough, S. and Penson, K.A.: 
Statistics on Graphs, Exponential Formula and Combinatorial Physics. 
In Proceedings of the \emph{3rd International Conference on Complex Systems and Applications
- ICCSA 2009}, Le Havre, France, (2009) 60–-63
%
\bibitem{Eil76}
Eilenberg, S.:
\emph{Automata, languages, and machines, vol. 1},
volume 59 of \emph{Pure and applied mathematics}.
Academic Press, 1976 
%
\bibitem{GP00}
Germain, C., Pallo, J.:
Langages rationnels d\'efinis avec une concat\'enation non-associative.
Theoretical Computer Science. \textbf{233} (2000) 217--231
%
\bibitem{Gra68}
Gr\"atzer, G.: 
\emph{Universal Algebra}.
D. Van Nostrand Company, Inc., 1968
%
\bibitem{Gud72}
Gudder, S.P.:
Partial algebraic structures associated with orthomodular posets.
Pacific Journal of Mathematics. \textbf{41}(3) (1972) 717--730
%
\bibitem{Hue80}
Huet, G.:
Confluent reductions: abstract properties and applications to term rewriting systems.
Journal of the ACM. \textbf{27}(4) (1980) 797--821
%
\bibitem{Hue81}
Huet, G.:
A Complete Proof of Correctness of the Knuth-Bendix Completion Algorithm.
J. Computer and. System Sciences \textbf{23} (1981) 11--21
%
\bibitem{Jan88}
Jantzen, M.:
\emph{Confluent string rewriting}, 
volume 14 of \emph{EATCS monographs on theoretical computer science}. 
Birkh\"auser, 1988 
%
\bibitem{KB70}
Knuth, D.E., Bendix, P.B.:
Simple word problems in universal algebras.
In Computational Problems in Abstract Algebra (Ed. J. Leech), (1970) 263-–297
%
\bibitem{Laf07}
Lafont, Y.:
Algebra and geometry of rewriting.
Applied Categorical Structures. {\bf 14}(4) (2007) 415--437
%
\bibitem{Lal90}
Lalement, R.:
\emph{Logique, r\'eduction, r\'esolution}.
\emph{\'Etudes et recherches en informatique}.
Masson, 1990
%
\bibitem{LeCh86}
Le Chenadec, P.:
\emph{Canonical forms in finitely presented algebras}.
\emph{Research Notes in Theoretical Computer Science}.
Pitman, John Wiley \& Sons, Inc., 1986
%
\bibitem{Ljap:eve}
Ljapin, E.S., Evseev, A.E.:
\emph{The theory of partial algebraic operations}, 
volume 414 of \emph{Mathematics and its applications}.
Kluwer Academic, 1997
%
\bibitem{McL98} Mac Lane, S.: 
\emph{Categories for the working mathematician (second ed.)}, 
volume 5 of \emph{Graduate Texts in Mathematics}.
Springer, 1997
%
\bibitem{New42}
Newman, M.H.A.:
On theories with a combinatorial definition of `equivalence'.
Annals of Mathematics. \textbf{43} (1942) 223--243
%
\bibitem{Seg73}
Segal, G.:
Configuration-spaces and iterated loop-spaces.
Invent. Math. 
\textbf{21}(3) (1973) 213--221 
%
\bibitem{Smi51}
Smith, P.A.:
The complex of a group relative to a set of generators. Part I.
Annals of Mathematics. \textbf{54}(2) (1951) 371--402
%
\bibitem{Sta63}
Stasheff, J.D.:
Homotopy Associativity of H-Spaces. I.
Transactions of the American Mathematical Society. \textbf{108}(2) (1963), 275--292
%
\bibitem{Tam71}
Tamari, D.:
Le probl\`eme de l'associativit\'e des mono\"ides et le probl\`eme des mots pour les demi-groupes; alg\`ebres partielles et cha\^ines \'el\'ementaires. 
S\'eminaire Dubreil-Pisot (Alg\`ebre et Th\'eorie des Nombres). \textbf{8} (1971) 1--15
%
\bibitem{Ter03}
Terese:
\emph{Term rewriting systems},
volume 55 of \emph{Cambridge tracts in theoretical computer science}.
Cambridge University Press, 2003	
%
\bibitem{Wil95}
Wilce, A.:
Partial Abelian semigroups.
Intern. J. Theor. Phys. \textbf{34}(8) (1995) 1807--1812 

\end{thebibliography}
\end{document}